\documentclass[pre,twocolumn,superscriptaddress,showpacs,preprintnumbers,amsmath,amssymb]{revtex4}
\usepackage{graphicx,epsfig}
\usepackage{dcolumn}
\usepackage{bm}
\usepackage{psfrag}
\usepackage{plain}
\usepackage{tabularx}
\usepackage[latin1]{inputenc}
\usepackage{amsmath}
\begin{document}

\title{A Hamiltonian approach for explosive percolation}

\author{A. A. Moreira}
\affiliation{Departamento de F\'{\i}sica, Universidade Federal
do Cear\'a, 60451-970 Fortaleza, Cear\'a, Brazil}

\author{E. A. Oliveira}
\affiliation{Departamento de F\'{\i}sica, Universidade Federal
do Cear\'a, 60451-970 Fortaleza, Cear\'a, Brazil}

\author{S. D. S. Reis}
\affiliation{Departamento de F\'{\i}sica, Universidade Federal
do Cear\'a, 60451-970 Fortaleza, Cear\'a, Brazil}

\author{H. J. Herrmann}
\affiliation{Departamento de F\'{\i}sica, Universidade Federal
do Cear\'a, 60451-970 Fortaleza, Cear\'a, Brazil}
\affiliation{Computational Physics, IfB, ETH-H\"onggerberg,
Schafmattstrasse 6, 8093 Z\"urich, Switzerland}

\author{J. S. Andrade Jr.}
\affiliation{Departamento de F\'{\i}sica, Universidade Federal
do Cear\'a, 60451-970 Fortaleza, Cear\'a, Brazil}

\begin{abstract}
We introduce a cluster growth process that provides a clear
connection between equilibrium statistical mechanics and an
explosive percolation model similar to the one recently proposed
by Achlioptas {\it et al.} [{\it{Science}} {\bf 323}, 1453
(2009)]. We show that the following two ingredients are
essential for obtaining an abrupt (first-order) transition in
the fraction of the system occupied by the largest cluster:
({\it i}) the size of all growing clusters should be kept
approximately the same, and ({\it ii}) the inclusion of merging
bonds (i.e., bonds connecting vertices in different clusters)
should dominate with respect to the redundant bonds (i.e., bonds
connecting vertices in the same cluster). Moreover, in the
extreme limit where only merging bonds are present, a complete
enumeration scheme based on tree-like graphs can be used to
obtain an exact solution of our model that displays a
first-order transition. Finally, the proposed mechanism can be
viewed as a generalization of standard percolation that
discloses an entirely new family of models with potential
application in growth and fragmentation processes of real
network systems.
\end{abstract}

\pacs{}

\maketitle

The second-order critical point of
percolation~\cite{Stauffer_1992,Sahimi_1994} has been
successfully used to describe a large variety of phenomena in
Nature, including the sol-gel transition~\cite{Brinker_1990},
or incipient flow through porous media~\cite{Stanley_1984}, as
well as epidemic spreading~\cite{Pastor_2001} and network
failure~\cite{Albert_2000,Newman_2002,Auto_2008,Moreira_2009}. A
long standing question of practical interest has been since, how
the transition could be made more abrupt and in the limit become
even of first-order. In other words, what ingredient must be
tuned in the basic model of random percolation to change the
order of the transition?


Recently Achlioptas {\it et al.}~\cite{Achlioptas_2009} proposed
a new mechanism on random graphs which they termed ``explosive
percolation'' that exhibits first-order phase transition. Their
process takes place in successive steps, with bonds being added
to the system in accordance to a selection rule. At each step, a
set of two unoccupied bonds are chosen randomly. From these two,
only the one with minimum weight becomes occupied. In
Ref.~\cite{Achlioptas_2009}, the weight is defined as the
product of the sizes of the clusters connected by this bond
(this is called ``product rule''). Importantly, if the bond
connects two sites that already belong to the same cluster, the
weight is proportional to the square of the cluster size. Since
unoccupied bonds connecting vertices in the largest cluster have
the largest possible weight, these bonds will become occupied
only if two of them are randomly chosen. Thus, this selection
rule hinders the inclusion of bonds connecting vertices that
already belong to the largest cluster. As a consequence, bonds
merging two smaller clusters will be selected more frequently,
resulting in the fast growth observed. Their model was then
implemented on a fully connected graph, however, it was shown
that the same effect takes place on 2D square
lattices~\cite{Ziff_2009} as well as  scale-free
networks~\cite{Cho_2009,Radicchi_2009}.

\begin{figure}[t]
\begin{center}
\includegraphics*[width=6.5cm]{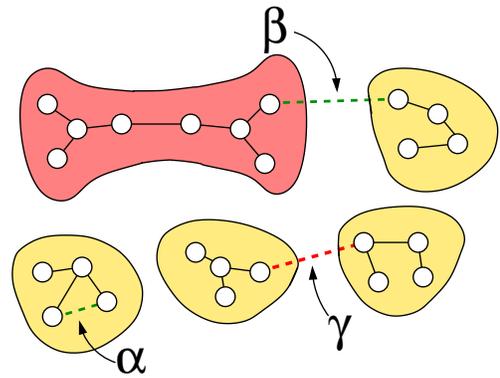}
\end{center}
\caption{
Two ingredients for explosive percolation. Here we show a possible
configuration for a growth process where, at each step, any unoccupied
bond can be introduced in the graph. For instance, in this figure we
show three bonds that could be added in the next step, namely,
$\alpha$, $\beta$, and $\gamma$. The two ingredients for obtaining a
sharp transition are the following: ({\it i}) bonds that keep the
clusters approximately at the same size are favored over bonds that
result in larger size discrepancies; and ({\it ii}) bonds that connect
vertices in the distinct clusters ({\it merging bonds}) are favored
over bonds that connect vertices in the same cluster ({\it redundant
bonds}). Thus, among the bonds indicated, $\alpha$ has the smallest
probability due to condition ({\it ii}), $\beta$ is not accepted due
to condition ({\it i}), and the most probable is the $\gamma$ bond.
}
\label{f.pic}
\end{figure}

In this letter we investigate what are the basic principles that lead
to the first-order phase transition observed in the explosive
percolation model. First we name {\it merging bonds} those edges that
connect vertices in distinct clusters, while {\it redundant bonds} are
edges connecting vertices in the same cluster. We show that two
conditions are necessary for obtaining a first-order transition in a
growth process where bonds are included one by one, namely, the
process has to favor the inclusion of bonds that keep all the clusters
at about the same size, and the process has to preclude the
introduction of redundant bonds, at least below  the critical
point. More precisely, merging bonds must be introduced with much
higher probability than redundant bonds. In Fig.~\ref{f.pic} we show a
pictorial description of these two ingredients.

In order to validate our hypothesis, we propose an extension of
the percolation model that describes a general growth process in
the space of graphs. For this, we define a Hamiltonian that
depends on the graph $G$ describing the network. The probability
of finding the system in a certain state $G$ will be given by
$P(G)=Z^{-1}\exp(-\beta{H(G)})$, where
$Z=\sum_{G}{\exp(-\beta{H(G)})}$.  A simple form for a
Hamiltonian that includes the two ingredients is
\begin{equation}
H(G)=\sum_{i \in \mathbf{C}}{s_{i}^2+\ell_{i} s_{i}^\alpha},
\label{eq.energ}
\end{equation}
where the sum is over the entire set of clusters $\mathbf{C}$, $s_{i}$
is the number of vertices in cluster $i$, and $\ell_{i}$ is the
number of redundant bonds added to this cluster. If the number of
bonds in the cluster is $b_{i}$, we have $\ell_{i}=1+b_{i}-s_{i}$. 
Note that each time one includes a redundant
bond, one also closes a new loop in the cluster, thus $\ell_{i}$ 
is also a measure for the number of loops in the cluster.

We can now simulate a process of cluster growth controlled by the
Hamiltonian of Eq.~(\ref{eq.energ}). This is performed by starting
with a network of $N$ vertices without bonds, so each vertex initially
belongs to a different cluster. At each step, a new bond can be placed
between any pair of vertices not yet connected. The probability of
including a particular bond $b$ is given by $\Pi_b\sim\exp(-\beta
\Delta{H_b})$, where $\Delta{H_b}$ is the energy change after including 
this bond. Such a growth model emulates equilibrium
configurations of graphs following the Eq.~(\ref{eq.energ}) and
having a given number of bonds $N_{b}$. However, since the
removal/rewiring of bonds is not considered during growth, this
corresponds to an out-of-equilibrium process. Consequently, some
differences should be expected between the observed results and
the actual thermal equilibrium.

For small values of $\alpha$, redundant bonds are favored over
merging bonds, while for large values of $\alpha$ the opposite
takes place.  Let us investigate the asymptotic behavior in the
two different scenarios. If redundant bonds are favored, one
might expect that a new merging bond will be included only after
the addition of all possible redundant bonds. Since clusters of
equal size minimize Eq.~(\ref{eq.energ}), we can assume that,
for low temperatures, all clusters have about the same size $S$,
so that fully connected sub-graphs with $S(S-1)/2$ bonds are
formed with $\ell=(S-1)(S-2)/2$. After adding the next bond, two
of these clusters shall merge to form a new largest cluster,
into which redundant bonds can be included. At this point we can
calculate the energy variation for a redundant bond,
$\Delta{H_r}=(2S)^\alpha$, and for a merging bond between pair
of clusters,
$\Delta{H_m}=2S^2+(S-1)(S-2)(2^\alpha-1)S^\alpha$. Surprisingly,
for any value of $\alpha$, in the asymptotic limit of very large
clusters, $S\to\infty$, merging bonds have higher energy
variation than redundant bonds, and the growth process with
fully connected clusters is stable.

\begin{figure}[t]
\begin{center}
\includegraphics*[width=7.5cm]{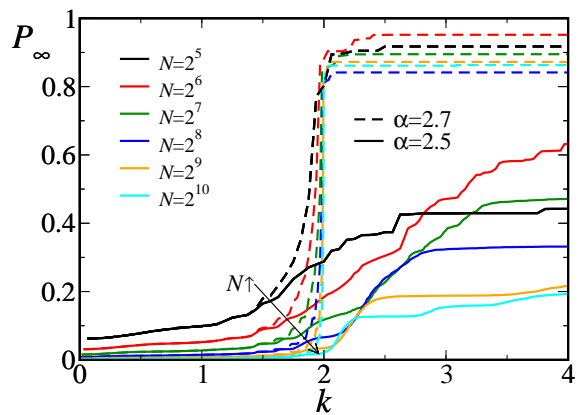}
\end{center}
\caption{
Transition to explosive percolation. When the process favors
redundant bonds, $\alpha=2.5$, the largest cluster follows a
slow continuous growth for the largest cluster. When the merging
bonds are favored, $\alpha=2.7$, the system displays an abrupt
transition around a critical connectivity $k=2$. In this case
the transition becomes sharper as the system size increases,
suggesting a first-order transition type of behavior. The inset
shows the total number of redundant bonds divided by the number
of nodes in the network, $N_{r}=N_{b}/N$. For $\alpha=2.7$,
redundant bonds are not included before $k=2$. In all
simulations, we use $\beta=1.0$ and take an average over $1000$
realizations of the growth process.
}
\label{f.2.6}
\end{figure}

The situation becomes quite different when the presence of merging
bonds is favored. As before, we use that all clusters have
approximately the same size $S$. However, without redundant bonds, the
clusters are all tree-like with exactly $S-1$ bonds, and $\ell=0$. At
this point, we have $\Delta{H_r}=S^\alpha$, and
$\Delta{H_m}=2S^2$. Thus, for large $S$, the inclusion of merging
bonds will lead to smaller energy variations, as long as
$\alpha>2$. We then conclude that, in the large cluster limit, $S\gg
1$, both scenarios are stable for $\alpha>2$.

The evolution of the system towards tree-like or fully connected
clusters is determined at the beginning of the growth process.
Considering that $S=3$ represents the minimal size necessary for
the inclusion of a redundant bond, we obtain
$\Delta{H_r}=3^\alpha$ and $\Delta{H_m}=2\times{3^2}=18$. Thus,
merging bonds become more probable when
$\alpha>\ln(18)/\ln(3)=2+\ln(2)/\ln(3)\approx{2.63}$, which
corresponds to a threshold condition above which the system
exhibits an abrupt transition. One should note that this is an
approximate result, since we do not account for fluctuations in
the cluster size distribution. However, as shown in
Fig.~\ref{f.2.6}, the results for $\alpha=2.5$ and $2.7$ indeed
confirm the change in behavior from a sharp transition for the
larger value of $\alpha$ to a slow continuous growth for the
smaller value. Note also that the threshold value for $\alpha$
is not universal and could be readily changed by adding a
multiplicative constant to any of the two terms constituting the
Hamiltonian of Eq.~(\ref{eq.energ}). In the inset of
Fig.~\ref{f.2.6}, we show the dependence of the fraction of
redundant bonds $N_{r}=N_{b}/N$ on the average connectivity of
the network $k$. As one can see, for $\alpha=2.7$, the inclusion
of redundant bonds is delayed up to $k\approx 2$, confirming
that the system is in the tree-like regime.

\begin{figure}[t]
\begin{center}
\includegraphics*[width=7.5cm]{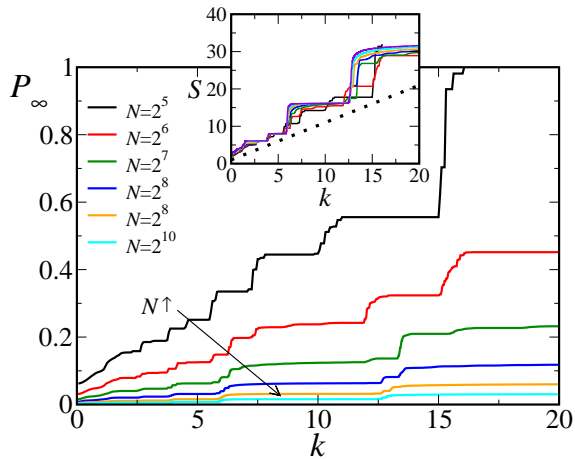}
\end{center}
\caption{
Growth process when redundant bonds are favored. Here we show
results for $\beta=1.0$ and $\alpha=2.0$. Since in this
situation merging bonds are less likely to be included, the
graph has to reach states where it splits in several fully
connected sub-graphs, before a new merging bond is
introduced. When the merging bond is included, a new and larger
cluster is created. This explains the presence of discontinuous
jumps in the size of the largest cluster.  Assuming that the
system consists of only fully connected clusters of the same
size, we obtain the dotted line shown in the inset,
$S=k+1$. This condition corresponds to the minimum bound for the
simulation results, that approximately follows this theoretical
prediction. Since the largest cluster $S$ is finite for any
finite connectivity $k$, the system does not display a
percolation transition.
}
\label{f.2.0}
\end{figure}

Let us examine in more detail the scenario for a small value of
$\alpha=2.0$. In Fig.~\ref{f.2.0} we show that the fraction
occupied by the largest cluster $P_\infty$ systematically
increases with the average connectivity $k$, with a growth rate
that decreases with system size $N$.  The inset of
Fig.~\ref{f.2.0} shows the same results, but for the size of the
largest cluster $S=N P_\infty$. One can see that $S$ follows
approximately a linear growth with the connectivity $k$. In this
scenario, a merging bond is expected to be placed only when all
clusters become saturated with redundant bonds.  If we now use
that all clusters have about the same size $S$, we obtain
$k=S-1$, which corresponds to the dotted line in the inset. The
deviations of the numerical results from this prediction should
be expected. In the growth model, the merging of two clusters
can only double the value of $S$, so that the values of $S$ at
the plateaus observed in the curves are approximately powers of
two.  However, we see that the curves always approach the dotted
line before doubling $S$. This linear growth for $P_\infty$ with
a slope that decays with the system size $N$, indicates that, in
the thermodynamic limit, this system does not undergo a
percolation transition, namely, $P_\infty=0$ for any finite
value of $k$.

\begin{figure}[t]
\begin{center}
\includegraphics*[width=7.5cm]{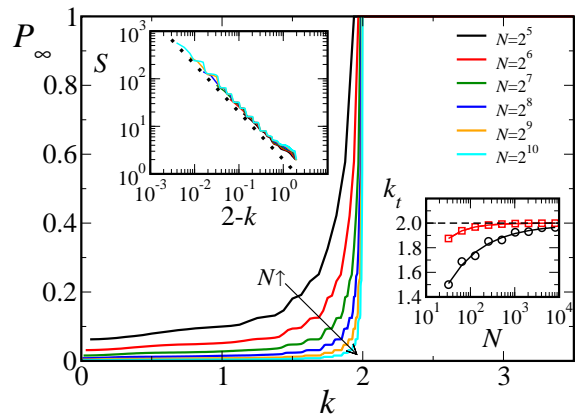}
\end{center}
\caption{
Growth process when merging bonds are favored. For $\beta=1.0$
and $\alpha=3.0$, the system does not include redundant bonds
and all clusters remain tree-like until the critical point
$k_{c}=2$ is reached. Supposing that the system comprises only
clusters of the same size $S$, we have that $S=2/(2-k)$ for the
case of trees. This relation works as a minimum bound for the
simulation results, as shown in the inset on the left. Thus we
have the critical condition $k_{c}=2$, where the size of the
largest cluster diverges to occupy the whole network. For
$k<k_{c}$ the largest cluster remains finite and its occupation
fraction $P_\infty$ goes to zero as the system size grows,
characterizing a typical first-order transition. The inset on
the right shows the threshold connectivity $k_t$ to obtain a
largest cluster greater than the square root of the system size,
$S>N^{1/2}$ (black circles), and greater than half the system
size, $S>N/2$ (red squares). The red line follows $k=2-4/N$, the
expected behavior for the connectivity where $S=N/2$. The black
line is a fit of the form $k=p_1+p_2\times N^{-p_3}$, with
$p_1=1.99\pm{0.03}$, $p_2=3.14\pm{0.02}$, and
$p_3=0.53\pm{0.05}$. In the limit $N\to\infty$ both curves
converge to $k\approx{2}$, that is in the thermodynamic limit we
observe at $k=2$ a discontinuous transition in the order
parameter from a vanishing fraction, $P_\infty\sim{N^{-1/2}}$,
to a finite fraction, $P_\infty=1/2$, confirming the approach to
a first-order transition.
}
\label{f.3.0}
\end{figure}

Figure~\ref{f.3.0} shows results for $\alpha=3.0$. Here we are in the
scenario where the clusters grow as loopless trees. In this case, the
system undergoes a transition that becomes sharper as the the number
of vertices $N$ increases. Again, if we assume that the system is
divided in trees of the same size $S$, we obtain $S=2/(2-k)$, as
indicated by the dotted line in the inset. As before, the size $S$
increases in steps due the out-of-equilibrium nature of the growth
process. Strikingly, the theoretical relation for the equilibrium
state still provides a consistent prediction for the lower bound of
the largest cluster size. Since $S$ remains finite for any $k<2$, and
at the critical point $k=2$ a tree that spans all the system is
formed, it follows that the order parameter $P_{\infty}$ displays a
first-order transition in the thermodynamic limit.

As already mentioned, this growth process bears some differences
with a thermal equilibrium state of graphs with the proposed
Hamiltonian Eq.~(\ref{eq.energ}) at low temperatures. In fact,
for $k \to 2$ there is always an energy gain in breaking large
trees in smaller highly connected graphs. One may then ask
whether the sharp transition observed in the simulations is just
a feature of the irreversible growth process or could be
reproduced in an equilibrium statistical framework. We now show
that in fact in the limit of large $\alpha$ we can obtain an
exact equilibrium solution that exhibits a first-order
transition.

If we impose that all clusters in the system are loopless trees,
$\alpha\to\infty$, it is possible to enumerate all possible ways in
which the network can be divided in a set of clusters of a given
size. Let $\Omega$ represent the number of ways that a fully connected
graph can be divided in trees with $n_{i}$ trees of size
$i=1,2,3...$ We then have
\begin{equation}
\Omega=N!\prod_s{\left(\frac{T_s}{s!}\right)^{n_s}\frac{1}{n_s!}},
\label{omega}
\end{equation}
where $N$ is the total number of vertices in the network, and
$T_s$ is the number of trees that span a fully connected graph
of size $s$, given by Cayley's formula,
$T_s=s^{s-2}$~\cite{Caley_1889}. Since all clusters are trees,
we can relate the number of clusters $N_{c}$ with the system
size $N$ and the average connectivity $k$ as $N_c=N(1-k/2)$.
Therefore, given a fixed value of $k$, the values of $n_s$ obey
the following two constraints: $\sum{n_s}=N_c$ and
$\sum{s{n_s}}=N$, where the sum is over all possible cluster
sizes $s$. In our generalized percolation model, we still need
to impose a fixed energy value, $\sum{E_s{n_s}}=E$, where
$E_s=s^2$ is the energy of a tree with size $s$. Using Lagrange
multipliers, $\eta$, $\lambda$, and $\beta$ to deal with each of
these constraints, we can find the cluster size distribution
that maximizes $\Omega$,
\begin{equation}
n_s=e^{\eta-\beta{s^2}-\lambda{s}}\frac{s^{s-2}}{s!}.
\label{ns}
\end{equation}

The critical condition takes place when the distribution
Eq.~(\ref{ns}) diverges. One can verify that, for $\beta=0$,
this happens when $\lambda=\lambda_c(\beta=0)=1$. The critical
connectivity can then be determined as
\begin{equation}
k=2\left(1-\frac{N_c}{N_s}\right)=
2\left(1-\frac{\sum{n_s}}{\sum{s n_s}}\right),
\label{kc}
\end{equation}
yielding $k_c=k(\lambda_c=1,\beta=0,\eta\to\infty)=1$. Note that, at
the critical condition $n_s\sim{s^{-5/2}}$, the fraction occupied by
the largest cluster follows $P_\infty=k-k_c$, thus reproducing the
known critical properties of the standard Erdos-Renyi
model~\cite{Erdos_1959}.

For $\beta>0$, the distribution always converges unless
$\lambda\to{-\infty}$. From Eq.~(\ref{kc}), we obtain that
$k_c=k(\lambda\to{-\infty},\beta,\eta\to\infty)=2$. For $k<k_c$
all clusters are finite trees, therefore occupying a vanishing
fraction of the network. At $k=k_c=2$, a giant tree spans the
entire network, characterizing a first-order transition. Of
course, this simple approach to the problem is only possible due
to the imposition of tree-like clusters. The general enumeration
of connected graphs with any number of redundant bonds is not a
simple task~\cite{Wright_1977}, and the cluster size
distribution in this generalized condition might be quite
different. However, at least in the situation where redundant
bonds are not present, explosive percolation can be duly
explained within the framework of equilibrium statistical
mechanics.

In summary, we have shown that two simple conditions, namely,
the absence of loops and the imposition of clusters of similar
sizes, are the only necessary ingredients for a percolation
process to display first-order transition in the size of the
infinite cluster as function of the average degree of the
network. We argue that both conditions are implicitly present in
the explosive percolation model proposed in
Ref.~\cite{Achlioptas_2009}. We emphasize that these conditions
are essentially non-local, namely, the probability of adding a
particular bond depends on the global structure of the graph.
Moreover, our model provides a simple connection between
explosive percolation and equilibrium statistical physics,
leading to a clear interpretation of the mechanisms behind this
growth process. Finally, other possibilities for the energy
function can also be investigated in different contexts,
revealing a whole new family of percolation-like models.

\end{document}